\documentclass[9pt,twocolumn,twoside]{opticajnl}
\journal{opticajournal} 

\setboolean{shortarticle}{true}
\usepackage[separate-uncertainty = true, multi-part-units=single]{siunitx}

\usepackage{lineno}

\title{Extremely high extinction ratio electro-optic modulator via frequency upconversion to visible wavelengths}

\author[1,*,\dag]{Alessandra Sabatti}
\author[1,\dag]{Jost Kellner}
\author[1]{Fabian Kaufmann}
\author[1]{Robert J. Chapman}
\author[1]{Giovanni Finco}
\author[1]{Tristan Kuttner}
\author[1]{Andreas Maeder}
\author[1]{Rachel Grange}

\affil[1]{Optical Nanomaterial Group, Institute for Quantum Electronics, Department of Physics, ETH Zurich, CH-8093 Zurich, Switzerland}

\affil[$\dag$]{These authors contributed equally to this work.}
\affil[*]{asabatti@phys.ethz.ch}

\begin{abstract}
Intensity modulators are fundamental components for integrated photonics. From near infrared to visible spectral ranges, they find applications in optical communication and quantum technologies. In particular, they are required for the control and manipulation of atomic systems such as atomic clocks and quantum computers.
Typical integrated electro-optic modulators operating at these wavelengths show high bandwidth and low voltage operation, but their extinction ratios are moderate. Here we present an integrated thin-film lithium niobate electro-optic modulator operating in the C-band, which uses a subsequent periodically poled waveguide to convert the modulated signal from \SI{1536}{\nm} to \SI{768}{\nm} using second harmonic generation. We demonstrate that the upconverted signal retains the characteristics of the modulated input signal, reaching a measured high bandwidth of \SI{40}{\GHz}. Due to the nature of the nonlinear process, it exhibits, with respect to the fundamental signal, a doubled extinction ratio of \SI{46}{\dB}, which is the highest recorded for near-infrared light on this platform.
\end{abstract}

\setboolean{displaycopyright}{false} 

\begin{document}

\maketitle

\section{Introduction}

The development of photonics in the spectral range spanning from the visible to the near infrared (NIR) is crucial for numerous applications, including free space communications \cite{visible_communication}, LiDAR \cite{lidar}, biosensing \cite{biosensing_visible}, augmented reality \cite{Augmented}, metrology and quantum technologies \cite{quantum_emitters}. Moreover, most of the atoms used in ion traps and cold atoms experiments, including calcium, caesium, and rubidium, have transitions in the visible or near-infrared spectrum ~\cite{trap_ion_home,trap_ion_clocks}. These experiments have applications in metrology~\cite{Rubidium_clock}, magnetometry with color centers in diamond~\cite{NVcenters}, single photons emitters~\cite{quantum_emitters}, and quantum computation~\cite{Silicon_visible, Integrated_quantum, trap_ion_er,Home}.The perspective of integrating these experiments on-chip would improve their scalability, reliability and efficiency.\\
One of the key components of integrated photonic circuits are intensity modulators. Besides being largely used in telecommunication, modulators in the visible and NIR range are especially suited for the excitation and control of atomic systems, where the most crucial parameters are large bandwidth and high extinction ratio (ER)~\cite{trap_ion_er, aln_mzi_er}.
Because of its strong electro-optic coefficient, wide transparency window, and low-loss waveguide architectures, lithium niobate on insulator (LNOI) is particularly well suited for realising such devices.\\
Visible modulators in LNOI have been demonstrated in an electro-optic (EO) Mach-Zehnder interferometer (MZI) configuration at a wavelength of \SI{738}{nm} and a bandwidth of \SI{35}{GHz}~\cite{sub-1_2023}, and at \SI{450}{nm} and a bandwidth of \SI{20}{GHz}~\cite{visible_mod}.
However, these devices only achieve ERs of \SI{21}{dB} and \SI{12}{dB} respectively, far below the requirements for applications such as control of trapped ions~\cite{trap_ion_er}.\\
In this work, we present a technique for modulating signals with very high ER. We realised a device that combines high-speed modulation in the C-band by means of an EO MZI, and second harmonic (SH) generation of the modulated signal in the NIR with a periodically poled lithium niobate (PPLN) waveguide \cite{ppln_shg}. The high-speed modulation is not performed directly on the visible light, but on the input light in the C-band, at \SI{1536}{nm}. The SH signal subsequently generated at \SI{768}{nm} inherits the modulation speed and the half-wave voltage value V\textsubscript{$\pi$} from the C-band signal, but it shows a doubled ER due to the quadratic dependence of the second harmonic power with respect to the fundamental. With our device, we measure a bandwidth of \SI{40}{GHz} and the increase in the ER by a factor of two, from \SI{23}{dB} in the fundamental to \SI{46}{dB} in the SH, which to the best of our knowledge is the highest reported value to date ~\cite{Desiatov:19}. 
This result demonstrates the possibility of realising extremely high ER devices working in the visible and NIR range, for example by frequency doubling the signal from devices that already show ER over \SI{50}{dB} in C-band ~\cite{Bragg_mod,HR_1550}. Additionally, the wavelength range is not constrained, and by changing the modulator and PPLN design, the same method can be adopted for various wavelengths.

\begin{figure*}[h!]
\centering\includegraphics[width=0.9\textwidth]{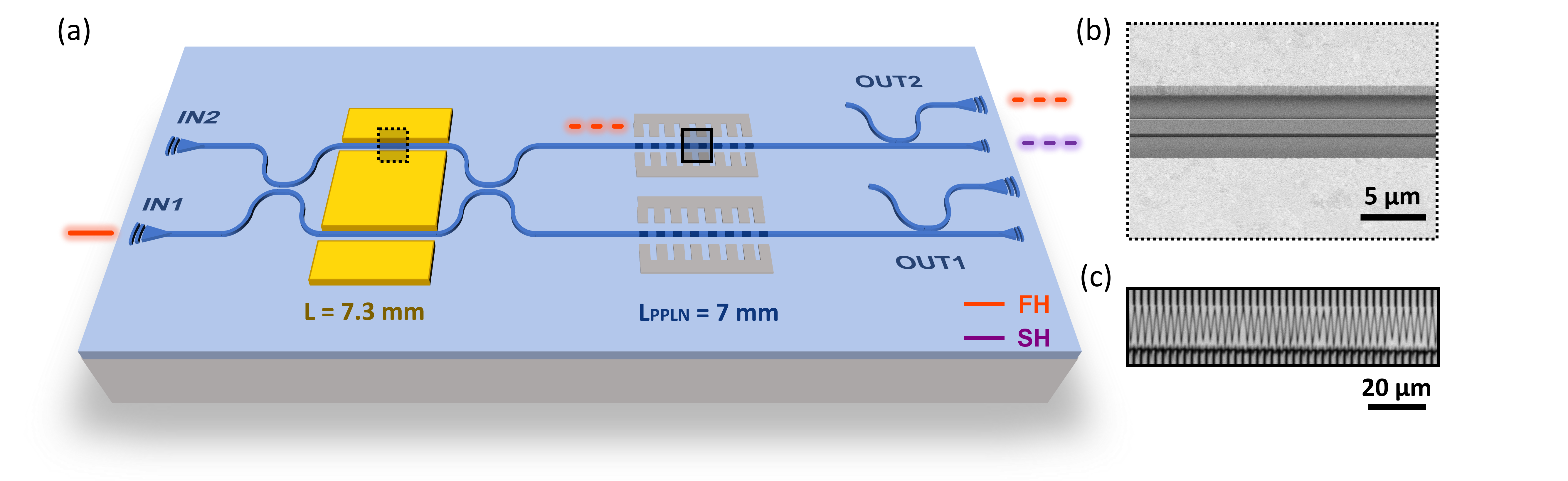}
\caption{Overview of the device. (a) Schematic of the circuit, showing input \SI{1550}{\nm} grating couplers, a Mach-Zehnder modulator, one PPLN waveguide at each output followed by wavelength division multiplexers for independent measurement of both wavelengths. The FH signal (red) enters the modulator and is transformed into an on-off signal. Afterwards, the signal generates an SH counterpart (purple) in the PPLN, where the SH signal inherits the properties of the modulated FH signal with double the ER. (b) SEM picture of an electro-optic tuning section (dotted area in a). (c) Two-photon microscope image of the PPLN (black square in a).}
\label{Fig1}
\end{figure*}

\section{Design}
Figure~\ref{Fig1}~(a) illustrates the design and working principle of the electro-optic modulator in combination with the PPLN waveguide for SH generation. 
For light coupling to and from the chip, grating couplers at the two design wavelengths of \SI{1550}{\nm} for the fundamental harmonic (FH) signal and \SI{775}{\nm} for the SH signal are used, with coupling efficiencies of \SI{-4}{\dB} and \SI{-7}{\dB} respectively. The waveguide top width is set to \SI{1.15}{\um} for low loss operation in the C-band. The device features two input ports with FH grating couplers. The light is split and recombined with 50:50 directional couplers in an MZI, generating the modulated FH signal (red) as shown in Fig.~\ref{Fig1}~(a). Each of the two outputs is routed to a \SI{7}{\mm} long PPLN region where the FH signal is converted to the SH signal (purple), inheriting from the modulated FH the V\textsubscript{$\pi$} and modulation speed with a doubled ER. To independently probe the FH light and the generated SH light, a wavelength division multiplexer (WDM) is placed after the PPLN region. Both FH and SH are coupled out of the chip using the respective grating couplers. The device can be operated using either of the inputs, and the output is chosen such that the MZI is in a cross-port configuration to maximize the ER. The presence of two output ports allows us to select the one with the PPLN waveguide with the quasi-phase matching condition closer to the designed FH of \SI{1550}{nm}.\\
We use coplanar waveguide electrodes in a push-pull configuration to modulate the light at the FH wavelength. Using a finite element simulation, the ideal electrode design for index and impedance matching is obtained. For high-frequency bandwidth, we match the group index n\textsubscript{g} at the FH wavelength to the radio frequency (RF) phase velocity index n\textsubscript{RF} at \SI{50}{\GHz}. The optimal configuration is found for a \SI{6}{\um} gap between ground and signal electrodes with a width of \SI{30}{\um} for the signal electrode. The value of n\textsubscript{g} and n\textsubscript{RF} simulated for our waveguide configuration are \SI{2.19}{} and \SI{2.17}{} respectively. \\
The PPLN period is chosen as \SI{2.48}{\um} for an etch depth of \SI{235}{\nm} and waveguide top-width of \SI{1.15}{\um}, with an average initial film thickness of \SI{312}{\nm}. The theoretically achievable nonlinear conversion efficiency is \SI{4280}{\% W^{-
1} cm^{-2}} \cite{Broadband_shg}, whereas, considering a \SI{7}{mm} long nonlinear region, the ideal conversion efficiency of the full device is \SI{2100}{\% W^{-1}}.

\section{Measurements}

\subsection{Optical characterisation}

\begin{figure}[ht!]
\centering\includegraphics[width=0.4\textwidth]{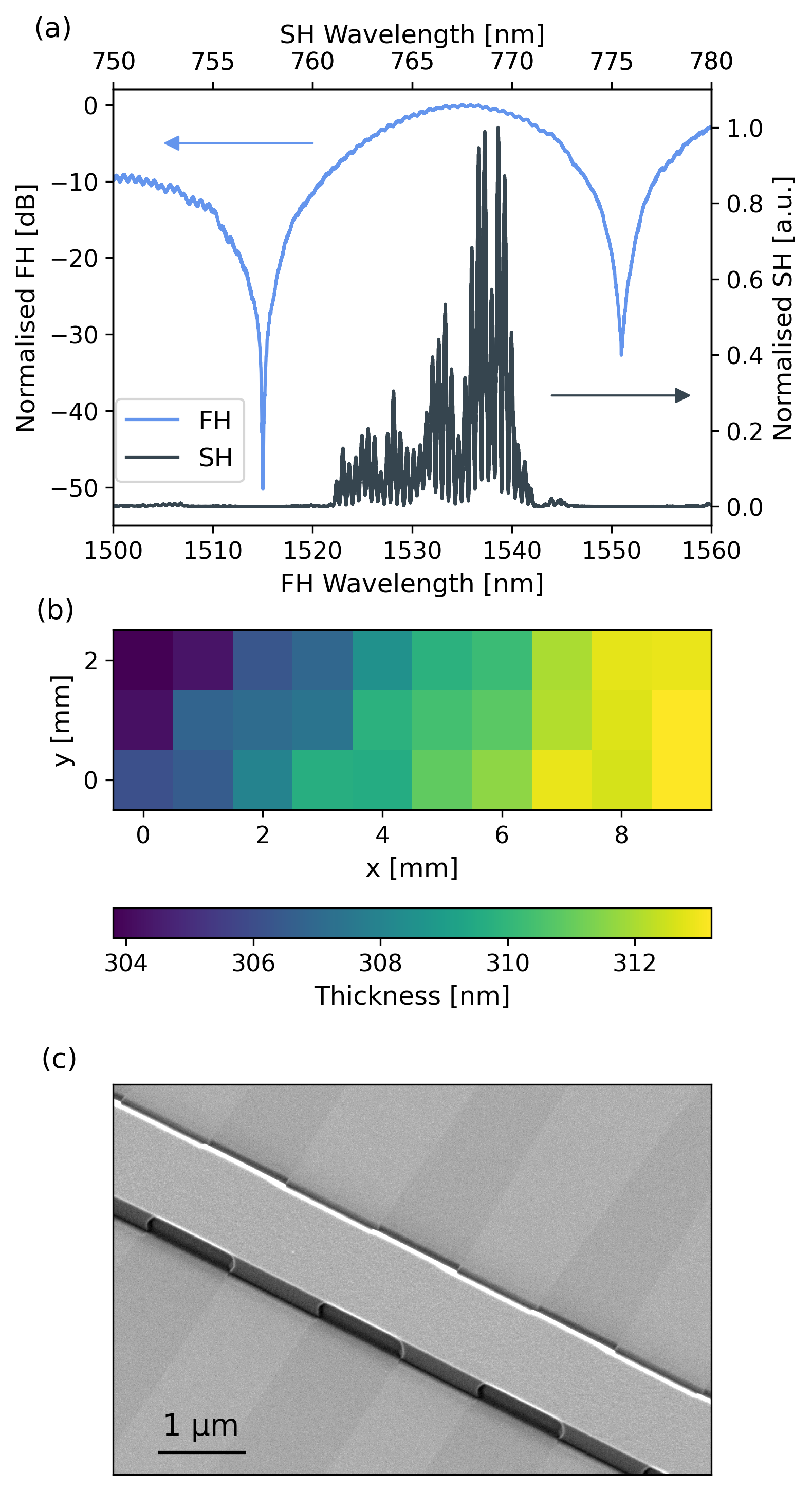}
\caption{Optical and morphological characterisation of the device. (a) Measurement of the fundamental (blue) and second harmonic spectra (black), defining the working wavelength of the device as \SI{1536}{nm}. The presents a broadened spectrum and reduced efficiency due to the film and waveguide properties. (b) Thickness of the lithium niobate thin film in the region of the chip corresponding to the device position,  measured with an optical reflectometer before etching the film. (c) SEM image of a periodically poled waveguide. A contrast between the lithium niobate regions with pristine crystal orientation (light grey) and the inverted domains (dark grey) is visible. In these sites the waveguide exhibits a profile variation that is attributed to the lateral etching rate variation with respect to the z-axis orientation of the lithium niobate crystal.}
\label{Fig2}
\end{figure}

We measure the spectral response of the full device at the FH and SH wavelength at OUT2 from IN1 as sketched in Fig.~\ref{Fig1}~(a) and shown in Fig.~\ref{Fig2}~(a). While the FH follows an MZI transmission function, the SH spectrum is expected to deviate from the theoretical sinc-squared function as it is generated by a spectrally non-constant FH signal. Besides this feature, the spectrum presents a larger bandwidth and lower peak efficiency than in the ideal case. This discrepancy can be explained by a variation of the lithium niobate thin film thickness in the region of the chip where the device is located, as shown in Fig.~\ref{Fig2}~(b). This difference, of about \SI{6}{nm} across the \SI{7}{mm} long PPLN region causes a broadening of the phase matching spectrum generated by a constant-period PPLN waveguide~\cite{Adapted_poling}. 
The conversion efficiency of the full device is measured as \SI{35}{\% W^{-1}}. This efficiency reduction is the outcome of the losses that the signal experiences in the MZI and also in the poled waveguide, which presents a lateral etching rate that is dependent on the crystal orientation, resulting in a corrugated shape~\cite{etch_orientation}. An SEM picture of a poled waveguide is shown in Fig.~\ref{Fig2}~(c), in which it is possible to distinguish the domains with opposite orientations, as they present a darker color, and the altered waveguide geometry that matches the domain profile.
Another reason for efficiency reduction is the variation in the phase matching condition, which reduces the effective length of the nonlinear interaction in the desired wavelength.
The FH operation wavelength for the characterisation of the modulator is set to \SI{1536}{nm}, resulting in a SH signal at \SI{768}{\nm}. 
\subsection{Electro-optical characterisation}
\begin{figure}[ht!]
\centering\includegraphics[width=0.4\textwidth]{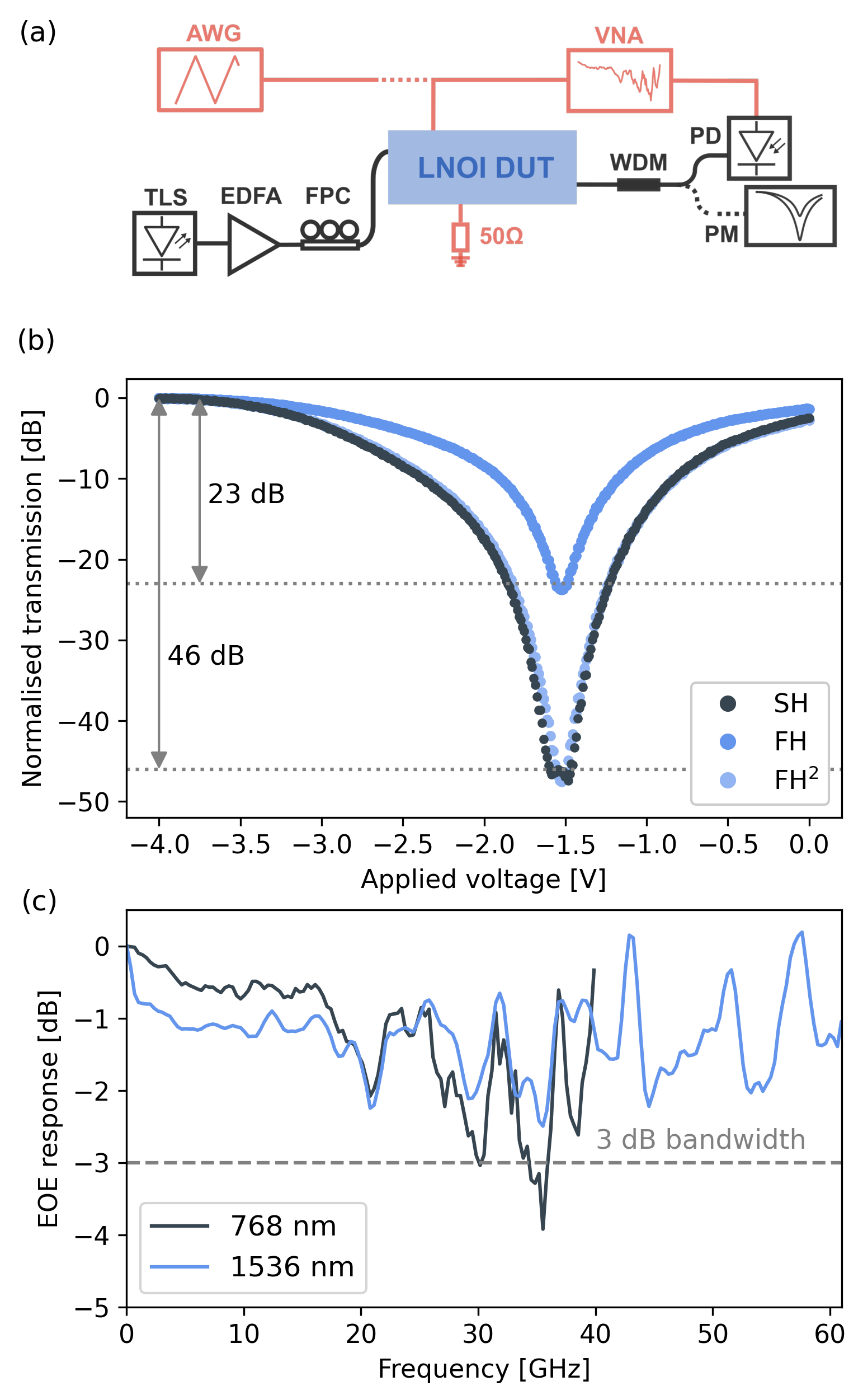}
\caption{Setup and electro-optic characterisation of the modulator with frequency upconversion. (a) Experimental setup illustration for low-frequency response including an arbitrary waveform generator (AWG) and power meter (PM) and high-speed measurements performed with a vector network analyser (VNA). (b) Measured slow EO response of the device featuring extinction rations for FH of \SI{23}{dB} and for SH of \SI{46}{dB}. The light blue dots indicate the squared FH signal for comparison.(c) Electro-optic (EO) response measurement from \SI{20}{MHz} to \SI{60}{GHz} for the fundamental harmonic (FH) and the second harmonic (SH), respectively.  }
\label{Fig3}
\end{figure}

The 7.3 mm long MZI modulator is optically characterised in the low-frequency range by applying a constant voltage to the electrodes via ground-signal-ground (GSG) probes (FormFactor ACP40) and acquiring the optical response on a high dynamic range power meter (Keysight N7744C). As illustrated in the measurement setup representation in Fig.~\ref{Fig3}~(a), an arbitrary waveform generator is used to apply the electrical signal. The FH signal from a tunable laser source (Keysight N7776C) is amplified in an erbium doped fibre amplifier (EDFA, KEOPSYS). An example of such a trace is presented in Fig.~\ref{Fig3}~(b). The $V_{\pi}$ for the FH signal and the upconverted signal is \SI{2.5}{V}, measured without termination of the modulator electrodes. The extinction ratio of the FH is \SI{23}{dB} and the one measured for the SH is \SI{46}{dB}, exactly double as predicted in the materials and methods section. For illustration purposes, we add the squared signal of the FH response which demonstrates an excellent overlap with the measured SH response.\\  
Figure~\ref{Fig3}~(c) presents the electro-optic response of our modulator. For this measurement, a vector network analyser (VNA, R\&S ZNA) is used together with a high-speed photodetector (PD, Newport 1014) as displayed in Fig.~\ref{Fig3}~(a). Additionally, we electrically terminate the modulator with 50~$\Omega$ to prevent undesired electrical back reflections and standing waves. We acquire the response from \SI{20}{MHz} to \SI{60}{GHz} and measure a bandwidth for the FH beyond our setup capabilities and a bandwidth of \SI{40}{GHz} for the SH. The bandwidth of the SH here is limited by the noise floor of the VNA and not by the working principle. Due to low conversion efficiencies of the PPLN region, the generated SH signal is weak and results in a \SI{33}{dB} lower RF power compared to the FH. Both measurements show the same features and we report a \SI{1}{\dB} difference between the FH and SH signal from \SI{20}{\MHz} to \SI{20}{\GHz} in EO response. Since the group index for \SI{300}{nm} LNOI at visible wavelengths is higher than the RF index, it is hard to index match them, while for the infrared wavelength (FH) it is naturally index matched \cite{sub-1_2023}. This gives the possibility to make monolithic high-bandwidth modulators for the NIR and visible wavelength range without the limits of index matching or the need for a hybrid platform and complex engineering of the electrodes \cite{100_GHZ_vis}. 

\section{Conclusion}
We have presented a LNOI device that modulates a fundamental harmonic (FH) signal in the telecom C-band at \SI{1536}{nm} with a bandwidth of \SI{60}{GHz} and upconverts the modulated FH signal in an \SI{7}{\mm} long PPLN region to the second harmonic (SH) signal at a wavelength of \SI{768}{nm} via the SHG process. The modulated FH shows a measured extinction ratio (ER) of \SI{23}{\dB}, while the ER of the SH signal is doubled due to the nature of the non-linear conversion, achieving a value of \SI{46}{\dB} with a bandwidth of \SI{40}{GHz}, limited by the photodiode range. The measured conversion efficiency from the entire device, featuring a \SI{7}{mm} long PPLN region, is \SI{35}{\% W^{-1}}. This much lower conversion efficiency is a consequence of the losses experienced by the FH along the poled region of device, and the thin film thickness variations, which cause a change in the phase matching condition in the device. Further engineering of the periodic poling could enable a highly efficient device with strong light outputs at both wavelengths. However, the generated second harmonic signal allows us to measure the modulated signal in the visible with enhanced ER with respect to the fundamental. The concept of the device shows potential in applications such as cold atom traps and atomic optical clocks, where the resulting increase of the ER for the SH signal by a factor of two, due to the nonlinearity, allows for well defined on- and off-states of the light used to manipulate and control atomic level transitions.

\section{Materials and methods}

\subsection{Fabrication}

The device is fabricated on a commercially available \SI{300}{\nm} thick, $5 \%$ magnesium oxide doped lithium niobate thin film on a \SI{2}{\um} thick silicon dioxide insulation layer on a silicon handle.
The periodic poling is realized through the application of a high voltage pulse of \SI{520}{\V} to \SI{100}{\nm} thick Cr electrodes, defined by electron beam lithography (EBL), electron beam evaporation and lift-off. Electrodes are subsequently removed and the optical layer is defined in a second EBL step using hydrogen silsesquioxane resist and etched into the thin-film using an optimised argon etching process \cite{FabPaper}. A cleaning step is then performed with KOH to remove the etching by-products and with buffered HF to remove the remaining mask. The RF electrodes are defined by direct laser writing lithography and standard lift-off process and electron beam evaporation of \SI{850}{\nm} of Au with a \SI{5}{nm} thick Cr adhesion layer.
\subsection{Theory}
The intensity response of a Mach-Zehnder intensity modulator can be described in the following way:
\begin{equation}
I_{FH} \left(\Delta\phi\right) = I_{FH,0} \left(r_1^2t_2^2+t_1^2r_2^2+2r_1t_1r_2t_2 \cos{\left(2\Delta\phi\right)}\right)
\label{Eq1}
\end{equation}
where $r_i,t_i$ are the reflection and transmission coefficients of the first and second directional coupler of the MZI. We can simplify the intensity response to $I_{FH} = I_{FH,0}\cos^2{\left(\Delta\phi\right)} + I_{FH,min}$. Depending on these coefficients the extinction ratio is defined by $ER_{FH} = 10 \log_{10}{\frac{I_{FH,min}}{I_{FH,0}}}$. Without loss of generality we can assume $I_{FH,0}=1$. If the modulated electric field from the modulator is sent to the PPLN region we can describe the intensity of the upconverted signal after the nonlinear interaction length $L$ 
\begin{equation}
I_{SH} \left(\Delta\phi\right) = \eta L^2 I_{FH}\left(\Delta\phi\right)^2
\end{equation}
where $\eta$ is the nonlinear interaction coefficient \cite{Fundamental_of_photonics}. With this knowledge we can now give the equation for the extinction ratio of the SH signal
\begin{equation}
ER_{SH} = 10 \log_{10}{\frac{I_{FH,min}^2}{I_{FH,0}^2}}=20 \log_{10}{\frac{I_{FH,min}}{I_{FH,0}}} = 2 ER_{FH}.
\label{Eq3}
\end{equation}
A similar calculation can be performed for the bandwidth of the electro-optic modulation to verify that the bandwidth of the SH signal is inherited from the FH.

\begin{backmatter}
\bmsection{Funding} This work was supported by ESA discovery idea Open channel project 4000137426 HEIDI and co-fund PhD fellowship. We acknowledge support from the Swiss National Science Foundation under the Ambizione Fellowship Program (Project Number 208707), Integrated Phase Logic Networks Sinergia grant (Project Number 206008), and LINIOS Bridge grant (194693). 

\bmsection{Acknowledgments} We acknowledge support for fabrication and characterization of our samples from the Scientific Center of Optical and Electron Microscopy ScopeM and from the cleanroom facilities FIRST and BRNC of ETH Zurich and IBM Ruschlikon.

\bmsection{Disclosures} The authors declare no conflicts of interest.

\bmsection{Data Availability} The data that support the findings of this study are available from the corresponding author upon reasonable request.
\smallskip


\end{backmatter}

\bibliography{bibliography}

\bibliographyfullrefs{bibliography}

\end{document}